\documentclass[aps,preprint]{revtex4}
\usepackage{amsfonts}
\usepackage{amsmath}
\usepackage{amssymb}
\usepackage{epsf}
\usepackage{epsfig}
\usepackage{graphicx}
\usepackage{rotating}
\usepackage{boxedminipage}

\begin{document}

\title{Local persistense and blocking in the two dimensional Blume-Capel
Model}
\author{Roberto da Silva }
\email{rdasilva@inf.ufrgs.br (corresponding author)}
\affiliation{Departamento de Inform\'{a}tica Te\'{o}rica, Instituto de Inform\'{a}tica,
Universidade Federal do Rio Grande do Sul. \\
Av. Bento Gon\c{c}alves, 9500, CEP 90570-051 Porto Alegre RS Brazil}
\author{S. R. Dahmen }
\email{dahmen@if.ufrgs.br}
\affiliation{Inst\'{\i}tuto de F\'{\i}sica, Universidade Federal do Rio Grande do Sul.\\
Av. Bento Gon\c{c}alves, 9500,CEP 90570-051 Porto Alegre RS
Brazil}

\begin{abstract}
In this letter we study the local persistence of the two--dimensional
Blume-- Capel Model by extension of the concept of Glauber dynamics. We
verify that for any value of the ratio $\alpha =D/J$ between anisotropy $D$
and exchange $J$ the persistence shows a power law behavior. In particular
for $\alpha <0$ we find a persistence exponent $\theta _{l}=0.2096(13)$,
\textit{i.e.} in the Ising universality class. For $\alpha >0$ ($\alpha \neq
1$) we observe the occurrence of blocking.
\end{abstract}

\maketitle

\setlength{\baselineskip}{0.7cm}

Among the many relevant quantities of interest in the modelling of
nonequilibrium dynamics of spin systems such as \emph{first time passage}
\cite{Feller}, which has been extensively discussed in the literature due to
the appearance of non--trivial exponents in the power law behavior of the
first return probability as function of time, another quantity of interest
is that of \emph{persistence}, \textit{i.e.} the characterization of the
time it takes for a particular spin \textit{not} to change its state from
its given $t=0$ configuration. Defining $P(t)$ as the probability that a
particular spin will not flip up to time $t$, at zero temperature
temperature and for the Ising and Potts models this quantity was shown to
behave as \cite{B.Derrida, Stauffer}
\begin{equation}
P(t)\sim t^{-\theta _{l}},  \label{eq:powerlaw}
\end{equation}%
where the \emph{persistence} exponent $\theta _{l}$ describes the
non-equilibrium relaxation of the system. This has been determined through
\textit{coarsening} simulations since one expects that the fraction of spins
which do not change up to $t$ represents a good estimate of $P(t)$. As a
consequence one may introduced a global version of the concept of
persistence through the quantity $P_{g}(t)$ which represents the probability
that the magnetization does not change its sign from its $t=0$ value, as
done in \cite{S.N.Majundar}. Recently one of the authors explored these
ideas to determine the associated $\theta _{g}$ exponent in the Blume Capel
Model, in particular its behavior at the critical and tricritical points
\cite{blumeglobal}. It was found that the exponent shows an abrupt change as
one goes from the critical points ($\theta _{g}\simeq 0.23$, Ising
universality class) to the tricritical one.

The purpose of this letter is to extend the Glauber dynamics to the Blume
Capel Model and to study the influence of the anisotropy on the exponent $%
\theta _{l}$. For the sake of clarity, we start with a brief overview of the
Ising model.

At $T=0$ the dynamics can be greatly affected by local \textit{blocking}
configurations, and the energy necessary to overcome them might be so high
as to render the system static. On the other hand, for $T\neq 0$ it becomes
difficult to define domains because one might not be able to distinguish
between true domains and spin flips due to thermal fluctuations. Nonetheless
for the Ising Model these difficulties can be overcome and a power law a
power law decay for the fraction of persistent spins was ascertained in the
whole low temperature phase \cite{Derrida}. The calculated values of the
exponent were $\theta _{l}=0.22$, $0.22$ and $0.29$ for $T=0$, $T=T_{c}/3$
and $T=2T_{c}/3$ respectively. Furthermore for $T>T_{c}$ an exponential
cutoff was verified. Indeed, using a natural definition of persistence, the
blocking effects is sensible to any temperature value, and power law only
happens exactly in $T=0$. The dynamics is implemented as follows: define the
excitation energy $\Delta E$ associated to the transition $\sigma
_{i}\rightarrow -\sigma _{i}$ as $\Delta E=-2J\sigma_{i}S_{i}$, with $S_{i}$
equal to the sum of nearest neighbors to $\sigma_{i}$. If $\Delta E < 0$
then the transition occurs with probability 1. For $\Delta E$ $=0$ it occurs
with probability $1/2$ and the transition does not occur if $\Delta E >0$.

To extend these ideas to the Blume-Capel Model we start out with the
Hamiltonian
\begin{equation}
H=-J\sum_{i}\sigma _{i}\sigma _{i+1}+D\sum_{i}\sigma _{i}^{2}
\end{equation}%
where $\sigma _{i}\in \{-1,0,1\}$, $J>0$ and $D$ represents an anisotropy.
As we shall see the behavior of the persistence exponent is strongly
dependent on the value of $\alpha =D/J$ yielding a nontrivial extension of
the Ising Model. To implement the Glauber dynamics in the ground state at ($%
T=0$) we consider the excitation energy (in units of $J$ ) for the
transition $\sigma _{i}\rightarrow \sigma _{f}$ as follows:
\begin{equation}
\Delta E/J=(\sigma _{f}-\sigma _{i})\left[ \alpha (\sigma _{f}+\sigma
_{i})-S_{i}\right]
\end{equation}%
with $S_{i}\in \lbrack -4,4]\cap \mathbb{Z}$. To extend the $T=0$ single
spin Glauber dynamics from a 2--state to a 3--state model is not
straightforward and some rules have to be introduced. Let the energy
differences in the transition from $\sigma _{i}$ to any other two spin
states $\sigma _{f}^{1}$ or $\sigma _{f}^{2}$ be represented by $\Delta
E_{1} $ and $\Delta E_{2}$ respectively. There are six possibilities to
consider:

\begin{enumerate}
\item If $\Delta E_{1,2}<$ $\Delta E_{2,1}<0$, then $\sigma _{i}\rightarrow
\sigma _{f}^{1,2}$;

\item If $\Delta E_{1,2}<0$ and $\Delta E_{2,1}\geq 0$, then $\sigma
_{i}\rightarrow \sigma _{f}^{1,2}$;

\item If $\Delta E_{1,2}=0$ and $\Delta E_{2,1}>0$, then $\sigma
_{i}\rightarrow \sigma _{f}^{1,2}$ with probability $=1/2$;

\item If $\Delta E_{1}=\Delta E_{2}<0$ $,$ $\sigma _{i}\rightarrow \sigma
_{f}^{1}$ or $\sigma _{i}\rightarrow \sigma _{f}^{2}$ with probability = $%
1/2 $;

\item If $\Delta E_{1}=\Delta E_{2}=0$ $,$ $\sigma _{i}$ goes to any of the
three states with probability = $1/3$;

\item If $\Delta E_{1}>0$ and $\Delta E_{2}>0$, the transition does not
occur.
\end{enumerate}

With these rules Monte Carlo simulations were performed at $T=0$ for an $%
L\times L$ square lattice with $L=160$. Since the sample number $N_s$ does
not play an important role because of the absence of significant statistical
fluctuations (see \textit{e.g.} \cite{Jain}) we used $N_{s}=200$ with $1000$
\ MC steps. After some exploratory simulations a total of $60$ different
values of $\alpha$ within $\alpha \in \lbrack -3,3]$ where chosen.

After the 300th MC step convergence to power laws are found, as can be seen
in figure 1. For $\alpha <0$ we obtained the exponent $\theta _{l}=0.214(3)$
by measuring directly the slope in the interval $[300,1000]$, with error
bars obtained using 5 different bins. A more precise estimate can be
obtained using the definition of the effective exponent (local slope):
\begin{equation}
\theta _{l}(t)=\frac{1}{\ln r}\ln \frac{\theta _{l}(t)}{\theta _{l}(t/r)}
\end{equation}%
\ \

\begin{figure}[tbph]
\centerline{\psfig{file=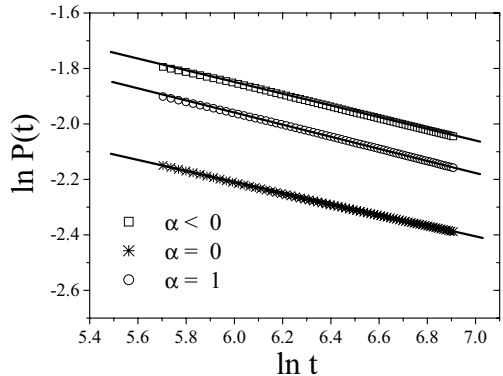,width=8cm}} \vspace*{8pt}
\caption{The numerical estimate of the probability $P(t)$ in (\protect\ref%
{eq:powerlaw}) as described in the text.}
\end{figure}

For $r=10$ and $t=1000$ we obtain $\theta _{l}=0.2096(13)$. The same power
law behavior was observed for $\alpha =0$ and $\alpha =1$, however with
slightly different exponents, as can be seen in table \ref{persloct=0bcalfa}%
.
\begin{table}[h]
\centering
\begin{tabular}{llll}
\hline\hline
\  & $\alpha <0$ & $\alpha =0$ & $\alpha =1$ \\ \hline\hline
$\theta _{l}$ & 0.2096(13) & 0.1964(34) & 0.1993(21) \\ \hline\hline
\end{tabular}%
\caption{Local persistence exponent values of Blume Capel Model}
\label{persloct=0bcalfa}
\end{table}
For positive values of $\alpha $ ($\neq 1$) the system shows blocking, as
can seen in figure 2. However a distinction must be made: in $\alpha \in $ $%
]0,1[$ the persistence has a fast decay and reaches a constant value $%
P(t\rightarrow \infty )\simeq 0.322$, while for $\alpha \in $ $]1,2]$ the
value is $P(t\rightarrow \infty )\simeq 0.246$. For $\alpha =2$, $%
P(t\rightarrow \infty )\simeq 0.287$) and when $\alpha >2$ there is full
blocking with $P(t\rightarrow \infty )$ frozen to the value $\simeq 0.333$
(initial mean fraction of spins is null). In $d=2$ and $T=0$ this is due to
the fact that for $\alpha >2$ the predominant phase is a configuration with
all spins zero.

\begin{figure}[th]
\centerline{\psfig{file=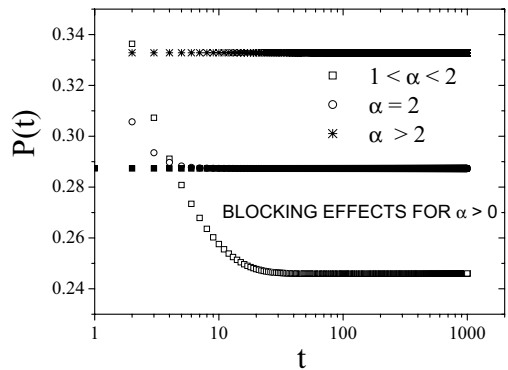,width=8cm}} \vspace*{8pt}
\caption{Blocking effects for positive values of $\protect\alpha $ ($\neq 1$%
). }
\end{figure}
Our values for $\alpha <0$ are in agreement with more recent results for the
Ising Model, namely
\begin{equation}
\theta _{l}=0.209(2)
\end{equation}%
\cite{jainflynm}. The point $\alpha =0$ separates two distinct regions: a
power law behavior for negative values of $\alpha $ and a blocking phase for
positive values of this ratio. At $\alpha =0$ we have a power law behavior
but with a different exponent.

Another interesting behavior is observed for $\alpha =1$, where a robust
power law separates two different blocking "phases". The point $\alpha =2$
also divides two regions of distinct blocking behavior: the $1<\alpha <2$
region and the full--blocking region $\alpha >2$.

To conclude, our results show that nontrivial behavior in stochastic
persistence can appear as one extends the Ising Model to allow for higher
"spins" and anisotropy effects as measured by our parameter $\alpha $. In
particular for $\alpha <0$ one has a pure 2--state Ising behavior; the same
is not observed for different values of $\alpha $, as discussed in the text.
It would be interesting to test these ideas for different stochastic systems
in the hope that there also nontrivial behavior might be found.

\section*{Acknowledgements}

S.R.D. would like to thank the Mathematical Physics Group at the University
of Queensland for their hospitality.

\end{document}